\documentclass[aps,prl,twocolumn]{revtex4}
\usepackage{newlfont}
\usepackage{amssymb}
\usepackage{amsfonts}
\usepackage{amsmath}
\usepackage{graphicx}
\usepackage{bm}
\newcommand{\ket}[1]{|{#1}\rangle}
\newcommand{\bra}[1]{\langle{#1}|}

\renewcommand{\t}[1]{\textrm{#1}}

\begin{document}

\title{Entanglement Enhances Security in Secret Sharing}

\author{Rafa{\l} Demkowicz-Dobrza\'nski\(^{1}\), Aditi Sen(De)\(^{3}\), Ujjwal Sen\(^{3}\),
Maciej Lewenstein\(^{2,3}\)}

\affiliation{\(^{1}\)Institute of Physics, Nicolaus Copernicus University,
ul. Grudziadzka 5, 87-100 Toru\'{n}, Poland\\
\(^{2}\)ICREA and \(^{3}\)ICFO-Institut de Ci\`encies Fot\`oniques, E-08860  Castelldefels (Barcelona), Spain
}

\begin{abstract}
We analyze tolerable quantum bit error rates in secret sharing protocols, 
and show that using entangled encoding states
is advantageous in the case when the eavesdropping attacks are local. We also provide a criterion for
security in
secret sharing -- a
parallel of the Csisz{\' a}r-K{\" o}rner criterion in single-receiver cryptography.
\end{abstract}

\maketitle

In the last few years,
the role of entanglement in different branches of physics
has been  studied extensively,
ranging from many-body physics \cite{adp,rmp} to  quantum information processing \cite{HHHH-RMP}.
In particular, the
qualities and thresholds of entanglement for optimal quantum communication performance have been found,
e.g. with regard to
teleportation \cite{HHH-BE-ebong-tele},  dense coding \cite{ref-dense-coding},
and  cryptography \cite{ref-crypto}.
The necessity of entanglement in quantum computation is still under investigation (see e.g. \cite{Vidal-natun}).
In a different context, there is an ongoing research on the behavior of
entanglement in e.g. quantum phase transitions \cite{rmp},
local cloning \cite{ref-cloning}, and local state distinguishing \cite{ref-distin}.




In this paper, we will investigate
the advantage of entanglement in the security of a quantum communication task,
known as secret sharing \cite{hillery1999, cleve},
which
is a communication scenario in which a sender Alice (\(A\)) wants to provide
a (classical) message to two
recipients (Bobs -- \(B_1, B_2\)),
%
%
in a way that each of the Bobs
individually knows nothing about the message, but they can recover its content once they cooperate.
In order to transmit a binary message string $\{a_i\}$, Alice can then
take a sequence of completely random bits $\{b_{1,i}\}$, send it to $B_1$, and at the same time send
a sequence $\{b_{2,i}\}=\{a_i \oplus b_{1,i}\}$ to
$B_2$, where $\oplus$ denotes addition modulo $2$.
Thus
$a_i = b_{1,i}\oplus  b_{2,i}$, assuring that the Bobs can recover the message if they cooperate, and yet
none of them can learn anything on the message of Alice on his own, since the sequences $\{b_{1,i}\}$,
$\{b_{2,i}\}$ are completely random.

An important issue is of course security, i.e. distributing the message in a way
that no third  (actually fourth!) party learns  about it. This can be achieved
using quantum cryptography (e.g. by the BB84 scheme \cite{eta-BB84}).
Alice simply has to establish secret random keys, independently, with both Bobs,
and use them as one-time pads to securely send bits in the way required by secret sharing.
We call this the BB\(84^{\otimes 2}\) protocol.
It has been argued \cite{hillery1999} that a more natural way of using quantum states in secret sharing is
to send entangled states to the Bobs,
and as a result, avoid establishing random keys with each of the Bobs separately,
by combining the quantum and classical
parts of secret sharing in a single protocol.
We call the protocol in \cite{hillery1999, cleve} as E4 (since it uses four entangled states).

In this paper, we consider security thresholds for both E4 and
BB\(84^{\otimes 2}\), i.e.
the highest quantum bit error rates (QBERs) below which one-way
distillation of secret key is possible. There are
three main results proven in the paper.
\emph{First}, we provide a criterion for security of secret sharing, for which one-way classical distillation
of secret key is possible between the sender and the receivers: the parallel of the
Csisz{\' a}r-K{\" o}rner criterion in (single-receiver, classical) cryptography \cite{csiszar1978}.
\emph{Secondly}, we find the \emph{optimal} quantum eavesdropping attacks on both E4 and
BB\(84^{\otimes 2}\), that are individual, without
quantum memory, and most importantly, \emph{local}. Note that an attack which acts by local operations
and classical communication (LOCC) on the particles sent through
the two channels (\(A \rightarrow B_1\) and
\(A \rightarrow B_2\)) is physically more relevant in this distributed receivers case.
We show that the threshold QBER for E4 is about 18.2 \% higher than that of BB\(84^{\otimes 2}\).
This shows, to our knowledge for the first time, that it is more secure to use entangled
encoding states in secret sharing.
\emph{Thirdly}, we provide an interesting general method for
dealing with local eavesdropping attacks.

\emph{The protocols.} In our setting, a secret sharing protocol can be characterized by
$\{\ket{\psi^{j,0}},\ket{\psi^{j,1}}, \sigma^{j,k}_1 \otimes \sigma^{j,k}_2\}$, where $j$ labels the different
encoding ``bases'' used, $\ket{\psi^{j,a}}$ are two-qubit states send by Alice to the Bobs if she uses basis $j$ and
wants to communicate the logical value $a$,
while $\sigma^{j,k}_1 \otimes \sigma^{j,k}_2$ is a set of
observables compatible with basis $j$ (so that if the corresponding measurement is performed by the Bobs, it allows them to recover
a proper logical bit of Alice).
In practice, \(B_1\) (\(B_2\)) randomly measures the observables \(\sigma^{j,k}_1\) (\(\sigma^{j,k}_2\))
on states received from Alice in each round.
After the transmission is completed, the Bobs announce the observables they have used in each round to Alice,
who, judging on whether this combination of observables is present in $\sigma^{j,k}_1 \otimes \sigma^{j,k}_2$
for the particular $j$ she had used in that round, tells the Bobs whether to keep or reject their measured results for that round
-- this is called the sifting phase.
The BB\(84^{\otimes 2}\) protocol is defined as
%
\begin{equation}
\begin{array}{c|c|c|c}
j & \ket{\psi^{j,0}} & \ket{\psi^{j,1}} & \sigma_1^{j,k} \otimes \sigma_2^{j,k} \\
\hline
1 & \ket{x_+} \ket{x_+},\  \ket{x_-} \ket{x_-}& \ket{x_+} \ket{x_-},\  \ket{x_-} \ket{x_+} &
\sigma_x \otimes \sigma_x \\
2 & \ket{x_+} \ket{y_+},\  \ket{x_-} \ket{y_-}& \ket{x_+} \ket{y_-},\  \ket{x_-} \ket{y_+} &
\sigma_x \otimes \sigma_y \\
3 &
\ket{y_+} \ket{x_+},\  \ket{y_-} \ket{x_-}& \ket{y_+} \ket{x_-},\  \ket{y_-} \ket{x_+} &
\sigma_y \otimes \sigma_x \\
4 &
\ket{y_+} \ket{y_+},\  \ket{y_-} \ket{y_-}& \ket{y_+} \ket{y_-},\  \ket{y_-} \ket{y_+} &
\sigma_y \otimes \sigma_y
\end{array}
\nonumber
\end{equation}
where $\ket{x_{\pm}}$ ($\ket{y_{\pm}}$) are eigenstates of the Pauli $\sigma_x$ ($\sigma_y$) matrix.
The fact that there are two states corresponding to a given $\ket{\psi^{j,a}}$ simply means that
each of them is sent randomly with probability $1/2$.
%
%
The E4 protocol \cite{hillery1999} (see also \cite{eita-GS}),
on the other hand, is defined as
\begin{equation}
\begin{array}{c|c|c|c}
j & \ket{\psi^{j,0}} & \ket{\psi^{j,1}} & \sigma_1^{j,k} \otimes \sigma_2^{j,k} \\
\hline
1 & \ket{\psi_+} & \ket{\psi_-} &
\sigma_x \otimes \sigma_x,\ -\sigma_y \otimes \sigma_y \\
2 & \ket{\psi_+^i}  &  \ket{\psi_-^i} &
\sigma_x \otimes \sigma_y, \ \phantom{-}\sigma_y \otimes \sigma_x,
\end{array}
\nonumber
\end{equation}
where \(\ket{\psi_\pm} = (\ket{00}\pm\ket{11})/\sqrt{2}\), \(\ket{\psi_\pm^i} = (\ket{00}\pm i\ket{11})/\sqrt{2}\), and
 \(\ket{0}\), \(\ket{1}\) are eigenstates of the Pauli \(\sigma_z\) operator.
%
The question is which of these protocols tolerates a higher QBER.
%
%
%
%
After the sifting phase, let the bits of Alice and the Bobs, obtained in a given set of rounds, be
 described by the probability distribution $p_{AB_1B_2}(a,b_1,b_2)$. The corresponding
QBER is
$\t{QBER}=\sum_{a,b_1,b_2} p_{AB_1B_2}(a,b_1,b_2) [1-\delta_{a,b_1\oplus b_2}]$.

\emph{Error correction and privacy amplification.} Knowing QBER, we want to perform an one-way error correction procedure, such that
all errors are corrected with arbitrarily high probability.
In standard (single-receiver) cryptography, error correction can be performed either from
the sender to the receiver, or vice-versa.
In secret sharing, there are two \emph{separated} receivers,
and each of them individually
has bits that are completely random. So there is no way for Alice to perform one-way error correction to Bobs --
whatever she sends to each of them individually, it will not be enough for them to correct errors, unless she sends the
total information which is of course not the solution we are after.

The only remaining option is that each of Bobs sends some information to Alice, judging on which she is able
to correct her bits $\{a_i\}$ in a way that for every $i$: $a_i = b_{1,i} \oplus b_{2,i}$.
Fortunately, this is indeed possible. We present here an idea how this can be achieved.
We will adapt for our needs, a standard method in classical
communication theory -- namely, that of random coding (see e.g. \cite{cover1991, nielsen2000, brassard1994}).
%
Let each of the three parties have \(n\) bits after the sifting phase. The error correction procedure
uses a random coding function $f:\{0,1  \}^n \to \{0,1\}^m$, known to all three parties (and the rest of the world),
where $m\leq n$ will be chosen later. This function assigns a random $m$-bit codeword to each of $2^n$ possible $n$-bit strings.
Error correction goes as follows:
$B_1$ and $B_2$ calculate $f(\{b_{1,i}\})$ and $f(\{b_{2,i}\})$ respectively,  and send their $m$-bit codewords to Alice.
After this, Alice looks for all $n$-bit sequences $\{b^\prime_{1,i}\}$, $\{b^\prime_{2,i}\}$ such that
$f(\{b^\prime_{1,i}\})=f(\{b_{1,i}\})$, $f(\{b^\prime_{2,i}\})=f(\{b_{2,i}\})$, and chooses a pair
$\{b^\prime_{1,i}\}$, $\{b^\prime_{2,i}\}$, for which the Hamming distance $\t{dist}(\{a_i\},
\{b^\prime_{2,i} \oplus b^\prime_{2,i}\})$ is minimal.
It can be shown
that in the limit $n \to \infty$,
this strategy is successful with arbitrarily high probability, provided
\begin{equation}
\label{eq:codelength}
m \geq n[1+h(\t{QBER})]/2,
\end{equation}
where $h(p)=-p \log_2p - (1-p)\log_2(1-p)$ is the binary entropy function.
This result is quite intuitive, since in
a standard bipartite error correction, the length of a codeword has to fulfill $m \geq n h(\t{QBER})$.
In secret sharing
however, the two Bobs together have to provide Alice with $nh(\t{QBER})+n$ bits. These
additional $n$ bits are needed, since
a sequence of one of Bobs taken separately is completely random for Alice. As a result each of Bobs
has to send a code of length given by Eq. (\ref{eq:codelength}).

After the error correction stage is completed, Alice and the Bobs need to perform privacy amplification,
in order to obtain a possibly shortened, but a completely secure key, on which an eavesdropper has no information.
Privacy amplification presents no additional difficulty in a secret sharing scenario, as compared to standard bipartite
cryptography, since its performance, in principle, requires no additional communication between Alice and the Bobs.
It is enough that all parties apply the same hashing function \cite{privacy} for shortening the key, and if there were no
errors, in the sense that for all $i$, $a_i=b_{1,i} \oplus b_{2,i}$, then there will be no errors in the shortened key.


\emph{LOCC attacks.} We will analyze security of the protocols with the following restrictions imposed on an eavesdropper:
(i)
Eavesdropper can perform only individual attacks;
(ii)
Individual attacks are LOCC operations with respect to partition of the encoding states
 between $B_1$ and $B_2$;
(iii)
Eavesdropper is not allowed any kind of quantum memory.
The restriction (i) means that an eavesdropper can
interact, in a given round, with only the quantum state send by Alice to Bobs in that round.
Restriction (ii) is at the heart of the problem we analyze, and is natural
in the distributed receivers scenario. Note here that if no LOCC condition is imposed,
then the security analyses of the two-receiver E4 and single-receiver BB84 protocols are isomorphic.
The justification of (iii) is based
on current technology limitations -- no long lasting quantum memory
has been developed so far.

Let the probability distribution \(p_{ABE}(a,b,e)\) describe
single-round bit values, \(a\) of Alice, \(b=b_1\oplus b_2\)
of the Bobs, and \(e\) of an eavesdropper, after the eavesdropper's attack and after the sifting stage is completed.
In single-receiver cryptography, the maximal one-way secret key distillation rate \(K\) is given by
the Csisz{\' a}r-K{\" o}rner criterion \cite{csiszar1978}:
$K = I(A:B)-\t{min}(I(A:E),I(B:E))$, where \(I(\phantom{A}:\phantom{A})\) is the mutual information
between the corresponding parties.
As discussed in previous paragraphs, error correction in secret sharing  can be performed only in one direction
 (from Bobs to Alice).
Thus the secret key distillation rate in case of secret sharing is $K = I(A:B)-I(B:E)$, which is
therefore
the
parallel of the
Csisz{\' a}r-K{\" o}rner criterion in (single-receiver) cryptography
\cite{csiszar1978}.

In order to analyze eavesdropping attacks, consider the state $\ket{\psi^{j,a}}$
being sent from Alice to Bobs.
Collaborating eavesdroppers \(E_1\), \(E_2\), acting on channels
conecting $A$ with $B_1$ and $B_2$ respectively, can perform an arbitrary LOCC operation $\mathcal{E}$
(completely positive trace-preserving LOCC map) to create
\(\rho^{j,a}_{B_1B_2E_1E_2} = \mathcal{E}(\ket{\psi^{j,a}}\bra{\psi^{j,a}})\).
The operation is LOCC with respect to the partition $B_1,E_1\  |\  B_2, E_2$.
Subsequently, $E_1$, $E_2$ perform an LOCC measurement on their subsystems in order to obtain information
about the
bit shared by Alice with Bobs, while sending possibly-perturbed subsystems $B_1$, $B_2$ to their legitimate
recipients. Without loosing generality,
 we can restrict this measurement to have only two possible outcomes ($0$ or $1$),
since only the value of a transmitted bit is of interest to the eavesdroppers.
Hence
we model the measurement by a two element positive operator valued
measurement (POVM):  $\Pi_{E_1E_2}(0)$, $\Pi_{E_1E_2}(1)$. Obviously $\Pi_{E_1E_2}(e) \geq 0$, and
$\Pi_{E_1E_2}(0) + \Pi_{E_1E_2}(1) = \openone_{E_1E_2}$,
but here we additionally impose the constraint that the measurements are
LOCC-based.

The probability distribution
\(p_{ABE}(a,b,e)\) is given by
\(\sum_j p(j,a) \t{Tr}
[ \mathcal{E}(\ket{\psi^{j,a}}\bra{\psi^{j,a}}) \Pi_{B_1B_2}(j,b) \otimes
\Pi_{E_1E_2}(e)
]\),
where $p(j,a)$ is the probability that $A$ sends the state $\ket{\psi^{j,a}}$ in a given round,
whereas  $\{\Pi_{B_1B_2}(j,b)\}$ is a POVM corresponding to Bobs' measurement in basis $j$ (compatible
with the state sent by Alice), where the sum  of their individual measured values, modulo $2$, is equal $b$:
$b=b_1 \oplus b_2$. Probability normalization condition reads $\Pi_{B_1B_2}(j,0)+\Pi_{B_1B_2}(j,1)=\openone_{B1B2}$.
We assume the convention that if one of Bobs (locally) performs a measurement characterized
by a Pauli matrix $\sigma_i$,  then he ascribes the bit value $0$ or $1$, once in a measurement he projects
on an eigenvector with eigenvalue $-1$ or $1$ respectively.
To make \(p_{ABE}(a,b,e)\) more revealing, we introduce non-trace-preserving completely positive operations
$\mathcal{E}_0$, $\mathcal{E}_1 :
\mathcal{H}^{\t{in}}_{B_1} \otimes \mathcal{H}^{\t{in}}_{B_2}
\mapsto  \mathcal{H}^{\t{out}}_{B_1} \otimes \mathcal{H}^{\t{out}}_{B_2}$
acting on the input and output Hilbert spaces of the Bobs, and defined as
\(\mathcal{E}_e(\varrho_{B_1B_2})= \t{Tr}_{E_1E_2}[\mathcal{E}(\varrho_{B_1B_2}) \Pi_{E_1E_2}(e) ]\).
\(\mathcal{E}_e\) represents
the disturbance experienced by a state transmitted to the Bobs, once the eavesdroppers have obtained a particular value $e$
in their measurement.
Notice that even though each operation $\mathcal{E}_e$ is not trace-preserving
the operation $\mathcal{E}_0 + \mathcal{E}_1$ is -- it corresponds to
a situation when one averages over the results of the eavesdroppers' measurement.
We can now write
\(p_{ABE}(a,b,e)=
\sum_j p(j,a) \t{Tr}
[ \mathcal{E}_e(\ket{\psi^{j,a}}\bra{\psi^{j,a}}) \Pi_{B_1B_2}(j,b)
]\).
%
It is now clear, that the eavesdropping strategy is completely defined by specifying
the two operations $\mathcal{E}_0$, $\mathcal{E}_1$, and for a given protocol
yields a joint probability distribution $p_{ABE}(a,b,e)$.


To calculate the QBER threshold,
one should now look for the highest value of QBER, for
which it is still possible to find eavesdropping LOCC operations $\mathcal{E}_e$, so that
the resulting probability distribution $p_{ABE}$ enjoys the property $I(A:B)=I(B:E)$.
%
%
%
%
Forgetting for the moment about the LOCC constraint,
the problem of finding the QBER threshold is
a semi-definite program.
To see this, let us denote
$\mathcal{H}^{\t{out}}= \mathcal{H}^{\t{out}}_{B_1} \otimes \mathcal{H}^{\t{out}}_{B_2} $,
$\mathcal{H}^{\t{in}}= \mathcal{H}^{\t{in}}_{B_1} \otimes \mathcal{H}^{\t{in}}_{B_2} $ and
recall the Jamio{\l}kowski isomorphism \cite{jamiolkowski1972}
between
completely positive
maps $\mathcal{E}_e$ and positive semi-definite operators
$P_{\mathcal{E}_e} \in \mathcal{L}(\mathcal{H}^{\t{out}} \otimes\mathcal{H}^{\t{in}})$:
\(P_{\mathcal{E}_e} = \mathcal{E}_e \otimes \mathcal{I} \left(\ket{\Psi^+}\bra{\Psi^+}\right)\),
where $\ket{\Psi^+} =\sum_{i=1}^{\dim \mathcal{H}^{\t{in}}} \ket{i} \otimes \ket{i}$ is an unnormalized maximally entangled state
in the space $\mathcal{H}^{\t{in}}\otimes \mathcal{H}^{\t{in}}$, and $\mathcal{I}$ is an identity operation
on
$\mathcal{H}^{\t{in}}$.
Hence our problem variables are
entries of two $16 \times 16$ matrices, which are required to be
positive semi-definite. Trace-preservation condition of
$\mathcal{E}_0 + \mathcal{E}_1$ translates
to a condition on positive operators:
$\t{Tr}_{\mathcal{H}^\t{out}}\left(P_{\mathcal{E}_0} + P_{\mathcal{E}_1} \right) = \openone_{\mathcal{H}^\t{in}}$.
This condition is obviously linear in the matrix elements of $P_{\mathcal{E}_e}$. Similarly,
$p_{ABE}$ is also linear,
and hence the
security condition is linear.
Finally, the QBER, which we want to maximize, is linear.
In order to deal with an LOCC constraint, we
will impose the weaker ``PPT constraint'': positivity after partial transposition of the
$P_{\mathcal{E}_e}$ operators -- we transpose subsystem
$\mathcal{H}^{\t{out}}_{B_2} \otimes \mathcal{H}^{\t{in}}_{B_2}$.
This is  a strictly necessary condition for LOCC
\cite{ref-nlwe}. However, we will show that the
optimal PPT maps are also LOCC.



\emph{Entangled vs. product encoding.}
We now present the solutions for maximal tolerable QBER for $\t{BB84}^{\otimes 2}$ and E4 protocols found
by solving the corresponding semi-definite programs,
using the SeDuMi package.
Although solving a semi-definite program provided us only with numerical solutions,
we were able to recognize their analytical form,
and hence all results presented are analytical.

For the $\t{BB84}^{\otimes 2}$ protocol,
the optimal
\(P_{\mathcal{E}^{\t{BB84}^{\otimes 2}}_0}\), in the computational basis,
\(=\frac{1}{18}\t{diag}[4,2,2,1,2,4,1,2,2,1,4,2,1,2,2,4]\) \(+\)
the 16x16 matrix \((\alpha_{i,j})\) whose only nonzero elements are
\(\alpha_{1,4}=\alpha_{5,8}=\alpha_{5,12}=\alpha_{9,12}=\alpha_{13,16}
=\alpha_{1,13}^{*}=\alpha_{2,14}^{*}=\alpha_{2,15}^{*}=\alpha_{3,14}^{*}=\alpha_{3,15}^{*}=\alpha_{4,16}^{*}=\alpha_{8,9}^{*} = i/9\),
\(\alpha_{1,16}=\alpha_{6,11}=2/9\),
\(\alpha_{2,3}=\alpha_{5,9}=\alpha_{6,7}=\alpha_{6,10}=\alpha_{7,11}=\alpha_{8,12}=\alpha_{10,11}=\alpha_{14,15}=1/9\),
\(\alpha_{7,10}=-\alpha_{4,13}=1/18\), and hermitian conjugates.
The optimal  $P_{\mathcal{E}_1^{\t{BB84}^{\otimes 2}}}$ has the same entries on the diagonal, and the anti-diagonal,
while the remaining ones
are multiplied by $-1$.
These optimal PPT maps will later on proven to be LOCC.
The optimal $\t{QBER}(\t{BB84}^{\otimes 2})=5/18 \approx 0.2778$.

\begin{figure}
  \includegraphics[width=0.48\textwidth,height=0.25\textheight]{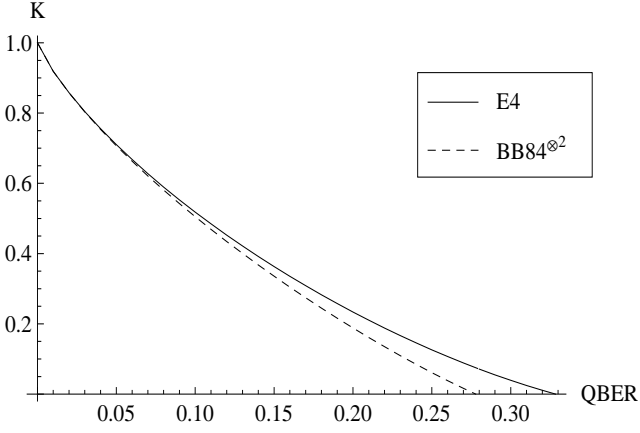}
  \caption{Maximal achievable secret-key rates \(K= I(A:B)-I(B:E)\) for E4 and
$\t{BB84}^{\otimes 2}$, against local attacks.
}\label{fig:compare}
\end{figure}

Moving now to the E4 protocol,
the optimal \(P_{\mathcal{E}_0^{\t{E4}}} =
\t{diag}[a,b,b,d,b,a,d,b,b,d,a,b,d,b,b,a]\) \(+\)
 the 16x16 matrix \((\beta_{i,j})\) whose only
nonzero entries are \(\beta_{1,4} = \beta_{1,13}^{*}= \beta_{4,16}^{*} = \beta_{13,16} =c\), \(\beta_{1,16} =a\),
\(\beta_{4,13} = f^*\), and the hermitian conjugates,
where $a=3-2\sqrt{2}$, $b=a /\sqrt{2}$, $c=b \exp(i \pi/4)$, $d=a/2$, $f=i d$.
The optimal $P_{\mathcal{E}_1^{\t{E4}}}$ is the same as $P_{\mathcal{E}_0^{\t{E4}}}$,
but with $c$ replaced by $-c$.
Again these optimal PPT maps will later on proven to be LOCC.
The optimal $\t{QBER}(\t{E4})=2(\sqrt{2}-5/4)\approx0.3284$. Interestingly therefore,
\(\t{QBER}(\t{E4})\) is about 18.2 \(\mathbb{\%}\) higher than \(\t{QBER}(\t{BB84}^{\otimes 2})\),
which indicates that indeed the protocol using
entangled states is more secure, in the case of LOCC eavesdropping.
In Fig.\ref{fig:compare}, we show the maximum achievable secret-key rates for the two protocols as a function of measured QBER.
It is clear that E4
is better not only because of its higher QBER threshold, but because of its higher key rate
for all QBER (see Fig.~\ref{fig:compare}, more details will be presented elsewhere \cite{ourfuture})


\emph{Explicit LOCC forms of the optimal attacks.}
We now show that the optimal attacks are separable.
We will subsequently show that
the attacks are actually LOCC.
%
%

Separability of the optimal attack for the BB\(84^{\otimes 2}\) case is evident once we write it in the form
(the procedure leading to this form will be presented elsewhere \cite{ourfuture})
\begin{equation}
\mathcal{E}_e^{\t{BB84}^{\otimes 2}}(\rho)=\sum_{\phi_1, \phi_2 \in \{0,\pi\}} K_{e,B_1}^{\phi_1,\phi_2}
 K_{e,B_2}^{\phi_1,\phi_2}
\rho K_{e,B_1}^{\phi_1,\phi_2 \dagger}
K_{e,B_2}^{\phi_1,\phi_2 \dagger}, \nonumber
\end{equation}
where the local Kraus operators $K_{e,B_1}^{\phi_1,\phi_2}$, $K_{e,B_2}^{\phi_1,\phi_2}$  are
\begin{align}
\frac{1}{\sqrt{6}}\left[\begin{array}{cc}
(-1)^e \sqrt{2} & \exp[i(\phi_1- \pi/4]\\
\exp[i(\phi_2+ \pi/4 )] & (-1)^e \sqrt{2}\exp[i(\phi_1+\phi_2)]
\end{array}\right], \nonumber
\\
\frac{1}{2 \sqrt{3}}\left[\begin{array}{cc}
 \sqrt{2} & \exp[-i(\phi_1+ \pi/4 )]\\
\exp[-i(\phi_2- \pi/4 )] &  \sqrt{2}\exp[i(\phi_1+\phi_2)]
\end{array} \right] \nonumber
\end{align}
respectively. Since $K_{e,B_2}$ does not depend on $e$
(equivalently, $K_{e,B_1}$ can also be chosen to be so),
we write it as
$K_{B_2}$.
The full operation $\mathcal{E}^{\t{BB84}^{\otimes 2}}=\mathcal{E}_0^{\t{BB84}^{\otimes 2}}+\mathcal{E}_1^{\t{BB84}^{\otimes 2}}$
can be written as
\(=\sum_{\phi_1, \phi_2 \in \{0,\pi\}} \openone \otimes K_{B_2}^{\phi_1,\phi_2}
( \sum_{e=0}^1
K_{e,B_1}^{\phi_1,\phi_2} \otimes \openone
\ \rho \ K_{e,B_1}^{\phi_1,\phi_2 \dagger} \otimes \openone
) \openone \otimes K_{B_2}^{\phi_1,\phi_2 \dagger}\),
which shows that it is
indeed LOCC,  since it can be realized as follows. First an operation given by the
four Kraus operators $K_{B_2}^{\phi_1, \phi_2}$
is performed on the second subsystem, and the measurement result $(\phi_1,\phi_2)$ is transmitted to the first subsystem.
For given values of $(\phi_1, \phi_2)$ received by the first subsystem, an operation using the
two Kraus operators $K_{0,B_1}^{\phi_1, \phi_2}$,
$K_{1,B_1}^{\phi_1, \phi_2}$ is performed on the first subsystem. This is a legitimate deterministic  LOCC operation
since
$\sum_{\phi_1,\phi_2 \in \{0,\pi\}} K_{B_2}^{\phi_1,\phi_2 \dagger}  K_{B_2}^{\phi_1,\phi_2}= \openone$, and for
every $(\phi_1, \phi_2)$, $\sum_{e=0}^1 K_{e,B_1}^{\phi_1, \phi_2 \dagger}K_{e,B_1}^{\phi_1, \phi_2}= \openone$.
Note that it requires only one-way classical communication.
Summing up, $\mathcal{E}_e^{\t{BB84}^{\otimes 2}}$ are separable trace-decreasing operations, such that
when added together, they form a trace-preserving LOCC operation $\mathcal{E}^{\t{BB84}^{\otimes 2}}$, and
hence
they can both be realized via LOCC.

In a similar way, we can show that the optimal PPT atacks on the E4 protocol are also LOCC.
Separable Kraus decompositions of $\mathcal{E}_e^{\t{E4}}$
read
\begin{equation}
\mathcal{E}_e^{\t{E4}}(\rho)=
\sum
K_{e,B_1}^{\phi_1,\phi_2,\phi_3}
K_{B_2}^{\phi_1,\phi_2,\phi_3}
\rho K_{e,B_1}^{\phi_1,\phi_2,\phi_3 \dagger}
K_{B_2}^{\phi_1,\phi_2,\phi_3 \dagger}, \nonumber
\end{equation}
where the sum runs over \(\phi_1, \phi_2, \phi_3 \in \{0,2\pi/3,4 \pi/3 \}\), and
\(K_{e,B_1}^{\phi_1,\phi_2,\phi_3}\), \(K_{e,B_2}^{\phi_1,\phi_2,\phi_3}\) are respectively
\begin{align}
\sqrt{1+\frac{1}{\sqrt{2}}}\left[\begin{array}{cc}
 (-1)^e 2^{1/4} & \exp(i\phi_1)\\
\exp(i\phi_2) &  (-1)^e 2^{1/4} \exp(i\phi_3)
\end{array}\right], \nonumber
\\
\frac{1}{\sqrt{27(1+\sqrt{2})}}\left[\begin{array}{cc}
 2^{1/4} & \exp[-i(\phi_1 + \pi/4)]\\
\exp[-i(\phi_2-\pi/4)] &  2^{1/4} \exp(-i\phi_3)
\end{array} \right]. \nonumber
\end{align}
%
Again we can write
the full operation $\mathcal{E}^{\t{E4}}=\mathcal{E}_0^{\t{E4}}+\mathcal{E}_1^{\t{E4}}$
as
\(
\sum_{\phi_1, \phi_2, \phi_3 \in \{0,2\pi/3, 4\pi/3 \}}
\openone \otimes K_{B_2}^{\phi_1,\phi_2,\phi_3}
\left(\sum_{e=0}^1  K_{e,B_1}^{\phi_1,\phi_2,\phi_3} \otimes \openone \ \rho \ K_{e,B_1}^{\phi_1,\phi_2,\phi_3\dagger} \otimes \openone \right)
\openone \otimes K_{B_2}^{\phi_1,\phi_2,\phi_3 \dagger},
\),
which shows that it is an LOCC,
since it can be realized by performing an operation on the second subsystem using the
27 Kraus operators
$K_{B_2}^{\phi_1,\phi_2,\phi_3 \dagger}$,
communicating the measurement result $(\phi_1, \phi_2,\phi_3)$ to the first subsystem, on which an
appropriate operation using the two Kraus operators $K_{e,B_1}^{\phi_1,\phi_2,\phi_3}$ ($e=0,1$) is performed.
Note that
$\sum_{\phi_1, \phi_2, \phi_3 \in \{0,2\pi/3, 4\pi/3 \}} K_{B_2}^{\phi_1,\phi_2, \phi_3 \dagger}
K_{B_2}^{\phi_1,\phi_2, \phi_3}= \openone$,
and for every $(\phi_1, \phi_2,\phi_3 )$, $\sum_{e=0}^1 K_{e,B_1}^{\phi_1, \phi_2, \phi_3  \dagger}K_{e,B_1}^{\phi_1, \phi_2,\phi_3}= \openone$.

\emph{Typical noise.} Judging the usefulness of the two protocols by comparing their QBER thresholds, may apriori
be not sensible from an experimental point of view, as in an experiment, we face noise caused by natural factors, as
well as
by the eavesdropper. Hence a relevant question is:
Which protocol allows a secure key transmission in presence of a higher level of noise, of
the type present in an experiment?
Consider a typical situation when we send the qubits via two fibers.
A usual model of noise here would be that each channel (fiber) is an isotropically depolarizing
channel -- and they are independent.
Given a channel
with a fixed level of depolarization, we ask: Can we securely extract some
secret key using either the E4 or the $\t{BB84}^{\otimes 2}$ protocol?
This may not be equivalent to comparing QBER thresholds, because different states are used in the two
protocols, which under the same noise level, may behave differently, and result in different
QBERs -- in particular it could happen that in such situation it might be
advantageous to apply a protocol with lower QBER threshold.
%
%
In this environment, however,
the QBERs for E4 and $\t{BB84}^{\otimes 2}$  depend in the same way on the depolarization parameter.
If an isotropically depolarizing qubit channel
acts as
 $\mathcal{D}(\rho)=(1-p)\rho + p\openone/2$, then the QBER caused by the $\mathcal{D}^{\otimes 2}$ channel is
$\t{QBER}=p(1-p/2)$ for \emph{both} the protocols. Comparing protocols using QBER thresholds as a figure of merit is legitimate
both from theoretical and practical point of view.

\emph{Summary.} We have for the first time shown that entanglement in the encoding states provide a
better security in secret sharing. The security was judged by calculating QBER threshold for secure communication,
under assumption of local individual quantum attacks
without quantum memory.  We have
found the optimal attacks in such scenario for the two paradigmatic protocols: one using product
states and the other using entangled ones for encoding. Further results include the parallel of the
Csisz\'{a}r-K\"{o}rner criterion for security in (single-receiver) cryptography in the
distributed-receivers case, and usefulness of the protocols in the presence of a depolarizing environment.

We acknowledge support from the Spanish MEC
(FIS-2005-04627, Consolider QOIT, Acciones Integradas,
\& Ram{\'o}n y
Cajal), ESF Program QUDEDIS, Euroquam FERMIX,
Polish Ministry of Science and Higher Education grant no.
1~P03B~011~29, EU IP SCALA, EU IP QAP.


\begin{thebibliography}{99}


\bibitem{adp} M. Lewenstein \emph{et al.},
Adv. Phys. \textbf{56}, 243 (2007).
\bibitem{rmp} L. Amico \emph{et al.},
to appear in Rev. Mod. Phys. (quant-ph/0703044).


\bibitem{HHHH-RMP} R. Horodecki \emph{et al.}, to appear in Rev. Mod. Phys. (quant-ph/0702225).

\bibitem{HHH-BE-ebong-tele} See e.g. C.H. Bennett \emph{et al.}, Phys. Rev. Lett.
\textbf{70}, 1895 (1993); P. Horodecki, M. Horodecki, and R. Horodecki,
Phys. Rev. A \textbf{60}, 1888 (1999).

\bibitem{ref-dense-coding} C.H. Bennett and S.J. Wiesner,
Phys. Rev. Lett. \textbf{69}, 2881 (1992).

\bibitem{ref-crypto} See e.g. A.K. Ekert, Phys. Rev. Lett. \textbf{67}, 661 (1991);
N. Gisin \emph{et al.}, Rev. Mod. Phys. \textbf{74}, 145 (2002);
K. Horodecki \emph{et al.}, \emph{ibid.} \textbf{94}, 160502 (2005).

\bibitem{Vidal-natun} A. Datta and  G. Vidal,
Phys. Rev. A \textbf{75}, 042310 (2007).


\bibitem{ref-cloning} See e.g. R. Demkowicz-Dobrza\'nski \emph{et al.},
Phys. Rev. A \textbf{73}, 032313 (2006).

\bibitem{ref-distin} See e.g. M. Hayashi \emph{et al.}, Phys. Rev. Lett.
\textbf{96}, 040501 (2006).





\bibitem{hillery1999}
M. \.Zukowski, A. Zeilinger, M. Horne, and H. Weinfurter,
 Acta Phys. Pol. \textbf{93}, 187 (1998);
  M. Hillery,  V. Bu\ifmmode \check{z}\else \v{z}\fi{}ek,   and A. Berthiaume,
Phys. Rev. A \textbf{59}, 1829
 (1999).

\bibitem{cleve} R. Cleve, D. Gottesman, and H.-K. Lo, Phys. Rev. Lett. \textbf{83},
 648 (1999);
 A. Karlsson, M. Koashi, and N. Imoto, Phys. Rev. A \textbf{59}, 162 (1999).

\bibitem{eta-BB84} C.H. Bennett and G. Brassard, in \emph{Proceedings of the
International Conference on Computers, Systems and Signal Processing,
Bangalore, India} (IEEE, NY (1984)).



\bibitem{csiszar1978} I. Csisz\'{a}r and J. K\"{o}rner,
IEEE Trans. Inf. Th.
 \textbf{IT-24},
339
 (1978)

\bibitem{privacy}
C.H Bennett, G. Brassard, C. Cr\'{e}peau, and U. Maurer, IEEE Trans. Inf. Theory, \textbf{41}, 1915 (1995).

\bibitem{eita-GS}  V. Scarani and N. Gisin, Phys. Rev. Lett. \textbf{87}, 117901 (2001);
Phys. Rev. A \textbf{65}, 012311 (2002);
A. Sen(De), U. Sen, and M. \.Zukowski,
Phys. Rev. A \textbf{68}, 032309 (2003); C. Schmid \emph{et al.},
Phys. Rev. Lett. \textbf{95}, 230505 (2005).


\bibitem{cover1991}
T.M. Cover and J.A. Thomas,
  \emph{Elements of Information Theory}
(Wiley, NJ (1991)).


\bibitem{brassard1994}G. Brassard and L. Salvail,
Adv. Cryptol. \textbf{765}, 410 (1994).


\bibitem{nielsen2000}
M.A. Nielsen and I.L. Chuang, \emph{Quantum Computing and Quantum Information} (CUP,
Cambridge (2000)).



\bibitem{jamiolkowski1972} A. Jamio{\l}kowski,
 Rep. Math. Phys.
 \textbf{3},
275 (1972).

\bibitem{ref-nlwe} P. Horodecki, Phys. Lett. A \textbf{232}, 333 (1997);
M. Horodecki, P. Horodecki, and R. Horodecki, Phys. Rev. Lett.
\textbf{80}, 5239 (1998);
C.H. Bennett \emph{et al.}, Phys. Rev. A \textbf{59}, 1070 (1999).

\bibitem{ourfuture} R. Demkowicz-Dobrza\'nski, A. Sen(De), U. Sen, and M. Lewenstein, in preparation.

\end{thebibliography}
\end{document}